\documentclass[11pt,a4paper]{article}
\usepackage{jcappub}
\usepackage{graphicx}
\usepackage{longtable}
\usepackage{amsmath}
\usepackage{amssymb}
\usepackage{subfigure}
\usepackage{hyperref}
\usepackage{dcolumn}
\usepackage{multirow}
\usepackage{cleveref}
\usepackage{bm}
\usepackage{color}

\newcommand{\be}{\begin{eqnarray}}
\newcommand{\ee}{\end{eqnarray}}
\newcommand{\mpl}{M_{\rm {pl}}}

\newcommand{\dd}{\, {\rm d}}
\newcommand{\gsim}{\;\mbox{\raisebox{-0.5ex}{$\stackrel{>}{\scriptstyle{\sim}}$}
}\;}

\def\eea{\end{eqnarray}}
\def\bea{\begin{eqnarray}}

\newcommand{\ccc}{_{\rm c}}

\newcommand{\gao}{\Gamma_{1,0}}

\newcommand{\U}{\Upsilon}

\newcommand{\s}{\sigma}

\newcommand{\nm}{{\mu\nu}}

\newcommand{\rpp}{r'}
\newcommand{\GN}{G_{\rm N}}

\newcommand{\dn}{\delta\nu}
\newcommand{\dl}{\delta\lambda}
\newcommand{\oo}{\mathcal{O}}

\newcommand{\lag}{\mathcal{L}}

\newcommand{\pdot}{v_0}

\newcommand{\goz}{\Gamma_{1,0}}

\allowdisplaybreaks
\begin{document}
\title{Stellar Pulsations in Beyond Horndeski Gravity Theories}
\author[a,b]{Jeremy Sakstein}
\emailAdd{sakstein@physics.upenn.edu}
\author[b]{Michael Kenna-Allison}
\emailAdd{mka1g13@soton.ac.uk}
\author[b]{Kazuya Koyama}
\emailAdd{kazuya.koyama@port.ac.uk}
\affiliation[a]{Center for Particle Cosmology, Department of Physics and Astronomy, University of Pennsylvania 209 S. 33rd St., Philadelphia, PA 19104, USA}
\affiliation[b]{Institute of Cosmology and Gravitation, University of Portsmouth, Portsmouth PO1 3FX, UK}

\abstract{Theories of gravity in the beyond Horndeski class recover the predictions of general relativity in the solar system whilst admitting novel cosmologies, including late-time de Sitter solutions in the absence of a cosmological constant. Deviations from Newton's law are predicted inside astrophysical bodies, which allow for falsifiable, smoking-gun tests of the theory. In this work we study the pulsations of stars by deriving and solving the wave equation governing linear adiabatic oscillations to find the modified period of pulsation. Using both semi-analytic and numerical models, we perform a preliminary survey of the stellar zoo in an attempt to identify the best candidate objects for testing the theory. Brown dwarfs and Cepheid stars are found to be particularly sensitive objects and we discuss the possibility of using both to test the theory.}
\maketitle

\section{Introduction}

Modern dark energy model building can be summarised by two words: \emph{modified gravity}. Despite the successes of general relativity (GR) in the century following its inception \cite{Will:2004nx}, the apparent late-time acceleration of the cosmic expansion requires either some form of exotic matter, or, as it is collectively known, dark energy \cite{Copeland:2006wr}, or is the result of some long-range infra-red modification of gravity \cite{Clifton:2011jh,Joyce:2014kja,Koyama:2015vza}\footnote{In many cases, the distinction between the two is not so clear \cite{Bull:2015stt}, for example, many scalar-tensor theories of gravity give rise to long-range forces on small scales but the cosmic expansion is driven by the dynamics of a rolling scalar field, just as in dark energy models such as quintessence.}. Perhaps the simplest explanation, a small ($\Lambda\sim$ meV$^4$) cosmological constant, is fraught with technical and theoretical obstructions since such a small value is not technically natural and requires an extreme fine-tuning \cite{Padilla:2015aaa}. This has led to a proliferation of dark energy/modified gravity models, many of which are scalar-tensor theories, and has motivated the search for interactions of ever-increasing complexity \cite{Horndeski:1974wa,Deffayet:2009wt,Deffayet:2011gz,Zumalacarregui:2013pma,Gleyzes:2014dya,Gleyzes:2014qga,Babichev:2015qma,Langlois:2015cwa,Deffayet:2015qwa,Langlois:2015skt,Crisostomi:2016tcp,Crisostomi:2016czh,Achour:2016rkg,Motohashi:2016ftl,BenAchour:2016fzp}. Typically, the requirement for a \emph{healthy} scalar-tensor theory is that it is free of the Ostrogradski ghost i.e. it propagates precisely three degrees of freedom\footnote{Often, one finds that specific models may have other forms of instabilities such as laplacian instabilities \cite{Kase:2014yya} or phantom behaviour \cite{Sakstein:2015jca}, in which case such models are discounted.}.  

With such a profusion of theories and models, one is inevitably led to the problem of distinguishing between one another observationally. Many of these theories fit into the parameterised post-Newtonian (PPN) framework for testing gravity in the solar system, in which case they are typically ruled out or are cosmologically uninteresting (see \cite{Ip:2015qsa} for example). Others can evade solar system tests altogether, either because they only exhibit novel effects in the strong field regime, as is the case with theories such as spontaneous scalarization \cite{Damour:1993hw,Chen:2015zmx,Minamitsuji:2016hkk}, or by employing non-linear screening mechanisms such as the chameleon \cite{Khoury:2003rn}, symmetron \cite{Hinterbichler:2010es}, or Vainshtein mechanisms \cite{Vainshtein:1972sx} (see \cite{Sakstein:2015oqa} for a brief introduction to these and see \cite{Sakstein:2014isa,Sakstein:2014nfa} for a discussion on linear screening mechanisms). 

Screening mechanisms are particularly interesting for cosmology because they typically admit novel accelerating solutions but their non-linear nature means they do not fall within the PPN framework, indeed, their modifications of the gravitational force in the solar system are highly suppressed compared with the Newtonian force. This has led several authors to propose a multitude of imaginative and innovative probes, which typically focus on laboratory or astrophysical probes (see \cite{Burrage:2016bwy} and references therein). 

One particularly interesting class of scalar-tensor theories that admit late-time de Sitter solutions without a cosmological constant are those known in the literature as \emph{beyond Horndeski} \cite{Gleyzes:2014dya,Gleyzes:2014qga}. These encompass a diverse set of scalar-tensor interactions and an expansive class of models. Modifications of gravity are screened in the solar system using the Vainshtein mechanism but said mechanism is ineffective inside astrophysical bodies of finite extent, resulting in modifications of Newton's law of the form \cite{Kobayashi:2014ida,Koyama:2015oma}
\begin{equation}\label{eq:mod1}
\frac{\dd\Phi}{\dd r}= \frac{\GN M}{r^2} +\frac{\Upsilon\GN}{4}\frac{\dd^2 M(r)}{\dd r^2}
\end{equation}
for spherically-symmetric objects. The dimensionless parameter $\U$ characterises deviations from GR and current constraints restrict its value to lie in the range $-0.44\le\Upsilon\le 0.027$ \cite{Koyama:2015oma,Sakstein:2015zoa,Sakstein:2015aac,Babichev:2016jom,Sakstein:2016oel}. It is related to the effective theory of linear cosmological perturbations in these theories \cite{Saito:2015fza,Sakstein:2015zoa,Sakstein:2016ggl}. Small-scale constraints on $\U$ therefore restrict the cosmology of beyond Horndeski theories on linear scales. This has motivated several authors to investigate astrophysical bodies, with several studies focusing on stars \cite{Davis:2011qf,Vikram:2014uza,Koyama:2015oma,Saito:2015fza,Sakstein:2015zoa,Sakstein:2015aac,Babichev:2016jom}. Hitherto, these studies have all concentrated on the equilibrium structure, but the oscillations about said equilibrium i.e. stellar pulsations have previously proved to be a powerful tool in the screening mechanism hunter's arsenal \cite{Jain:2012tn,Sakstein:2013pda,Sakstein:2014nfa}. These are the subject of this work.

The main body of this paper is devoted to stellar pulsations in beyond Horndeski theories, but we present the specific model and derive our important results in Appendix \ref{sec:model} and \ref{sec:deriv} respectively. This is done partly for the benefit of the unfamiliar reader and partly due to the length and technical nature of the derivations. In section \ref{sec:pulsations} we present the equation governing linear adiabatic oscillations of stars in beyond Horndeski theories and discuss its properties. We identify a new potential instability whereby there are growing modes when $\U\gsim49/6$ but it is unlikely that this is exhibited in physically sensible stellar models since values of $\U$ this large are ruled out observationally. Next, we perform a thorough investigation into the pulsation frequencies of the stellar zoo in an attempt to identify the best objects for testing these theories. We numerically solve the wave equation for both semi-analytic (section \ref{sec:semian}) and numerical models (section \ref{sec:num}) and report the fundamental frequencies, as well as the first and second harmonics. We identify Cepheid stars as particularly promising candidate objects and discuss the prospects for using pulsating stars to constrain $\U$ in section \ref{sec:concs}.

\section{Stellar Pulsations in Beyond Horndeski Theories}\label{sec:pulsations}

The equilibrium structure of stars is governed by the hydrostatic equilibrium equation
\begin{equation}\label{eq:hse}
\frac{\dd P}{\dd r}=-\rho\frac{\dd\Phi}{\dd r} = -\frac{\GN M\rho}{r^2}-\frac{\U\GN\rho}{4}\frac{\dd^2 M}{\dd r^2},
\end{equation}
where the last term represents the beyond Horndeski modification to the GR equation. To move beyond equilibrium we introduce the Lagrangian displacement perturbation
\begin{equation}
\vec{\delta r} = \delta r \vec{\hat{r}},
\end{equation}
where we have specialised to the case of perturbations along the radial direction only\footnote{The inclusion of transverse perturbations leads to the appearance of non-radial modes. Whereas such modes may be interesting for testing gravity because the Sun oscillates in several such modes, we specialise to purely radial modes in this paper. As we will see in Appendix \ref{sec:deriv}, the complex nature of beyond Horndeski theories requires us to do a fully relativistic calculation (including cosmological asymptotics) in order to find the non-relativistic limit. The inclusion of non-radial perturbations would likely make this problem intractable.}. The time-dependence of this perturbation is governed by the momentum equation
\begin{equation}\label{eq:eqhse}
\frac{\partial^2\delta r}{\partial^2t^2}=-\frac{1}{\rho}\frac{\partial P}{\partial r}-\frac{\dd\Phi}{\dd r}.
\end{equation}
The only gravitational contribution to the motion is from the Newtonian potential $\Phi$, which we derive in Appendix \ref{sec:deriv}. Treating $\delta r$ as a small perturbation and linearising the pressure and the gravitational potential, the resultant equations can be combined (see Appendix \ref{sec:deriv}) to find a single wave equation for the dimensionless displacement $\zeta(r)=\delta r/r = \xi(r) \exp(i\omega t)$:
\begin{equation}\label{eq:MLAWE}
\frac{\dd}{\dd r}\left[r^4\left(\Gamma_{1,0}P_0+\Upsilon\pi\GN r^2\rho_0^2\right)\frac{\dd\xi}{\dd r}\right]+r^3\xi\frac{\dd}{\dd r}\left[\left(3\goz-4\right)P_0\right]+r^4\rho_0\omega^2W(r,\Upsilon)\xi,
\end{equation}
where
\begin{align}
\Gamma_{1,0}&=\left(\frac{\partial \ln P}{\dd\ln\rho}\right)_{\rm adiabatic},\\
W(r,\Upsilon)&=1+\frac{\Upsilon\pi\GN r\rho_0^2(P_0'+\Upsilon\pi\GN r\rho_0[2\rho_0+r\rho_0'])}{2(P_0'+\pi\GN r\rho_0[\Upsilon r\rho_0'+(2\Upsilon-1)\rho_0])^2}\nonumber\\
&=1-\frac{ \Upsilon\pi   r^3\rho_0 M}{2 \left(M+\pi  r^3\rho_0\right)^2},\label{eq:weight2}
\end{align}
and the second equality is found by applying the equilibrium equation \eqref{eq:eqhse}. Throughout this work we use subscript zeros to refer to equilibrium quantities i.e. those that are determined by solving the hydrostatic equilibrium equation. In GR, equation \eqref{eq:MLAWE} is known as the \emph{Linear Adiabatic Wave Equation} and we refer to it as such here. It is a Sturm-Liouville eigenvalue equation for the pulsation frequency $\omega=2\pi/T$ with weight function $W(r,\Upsilon)$ and reduces to the GR equation when $\U=0$. Given an equilibrium stellar configuration $\{P_0(r),\,\rho_0(r),\,\gao(r)\}$, the eigenvalues are determined by finding mode functions that satisfy the appropriate boundary conditions at the centre and radius (which we discuss below). It is interesting to note that the terms in equation \eqref{eq:MLAWE} that are modified by beyond Horndeski theories, which, in some sense, are generalisations of galileon theories\footnote{Indeed, the specific model we consider in appendix \ref{sec:model} are covariantisations of the quartic galileon.}, are the kinetic terms (including the weight function, which multiplies a terms that arises due to time-derivatives) whereas chameleon theories modify the mass-like terms (proportional to $\xi$ with no factors of $\omega$) \cite{Sakstein:2013pda}.

Typically, one cannot find analytic solutions of the LAWE except in the most trivial and unphysical cases and so we solve it numerically in the next two sections. One can, however, deduce many important properties pertaining to stellar stability, which we discuss below.

\subsection{Boundary Conditions}

Spherical symmetry requires that only mode functions where $\xi(r)$ is non-singular at the centre are physical, which selects solutions with $\dd\xi/\dd r = 0$ at $r=0$. The boundary condition at the stellar surface is more complicated, and typically requires one to couple the LAWE to atmospheric models, but, for the lowest few modes, a good approximation is that the pressure is zero at the surface. Using the LAWE, this gives the second requisite boundary condition
\begin{equation}
\left(1+\frac{\goz-1}{\goz}\alpha\right)\frac{d\xi}{dr}\Big|_R = \frac{1}{\goz R}\left[\frac{\omega^2R^3}{GM}-\left(3\goz-4\right)\left(1+\alpha\right)\right] \xi;\quad\alpha=\frac{\pi\Upsilon  R^4}{M} \frac{d\rho_0}{dr},
\end{equation}
where $\rho_0'$ is evaluated at the surface and is not necessarily zero \cite{Koyama:2015oma}.

\subsection{Stability}

The LAWE is a Sturm-Liouville eigenvalue problem of the form $\hat{\mathcal{L}}\xi=\tilde{W}(r,\Upsilon)\omega^2\xi$ with weight function $\tilde{W}(r,\Upsilon)=-r^4\rho_0W(r,\Upsilon)$. As such, one can construct the functional 
\begin{equation}
F[\chi]=\frac{\int\dd r\, \chi^\dagger\hat{\mathcal{L}}\chi }{\int \dd r\, \chi^\dagger\chi\tilde{W}},
\end{equation}
which has the property that the smallest eigenvalue $\omega_0$ satisfies $\omega_0^2\le F[\chi]$ for any choice of test function $\chi$\footnote{Equality holds when one chooses $\chi=\xi_0$, the eigenfunction corresponding to $\omega_0$ i.e. $\hat{\mathcal{L}}\xi_0=\omega_0^2\tilde{r}(r)\xi_0$.}. If $F[\chi]<0$ there are growing modes because $\omega_0^2<0$ and therefore the star is unstable. Using the LAWE and setting $\chi$ to be a constant we find 
\begin{equation}
\omega_0^2<\frac{\int\dd r\, 3r^2P_0\left(3\goz-4\right)}{\int\dd r \, r^4\rho_0 W(r,\Upsilon)}.
\end{equation}
Treating $\goz$ as constant, this leads to the well-known result in GR that stars are unstable when $\goz<4/3$ i.e. at the onset of fully-relativistic motion\footnote{There is a GR correction to this Newtonian result that is negligible for non-relativistic objects such as those we study here \cite{Chandrasekhar:1964zza}.}. Looking at equation \eqref{eq:weight2}, one can see that there is a second potential instability in beyond Horndeski theories because one can have $W(r,\Upsilon)<0$ for large enough $\U>0$. We can estimate the critical value of $\U$ for the onset of this instability by taking a star of constant density so that $\rho_0=3M/4\pi R^3$. In this case, one finds that the instability occurs for $\U\gsim49/6$ taking $r\approx R$. A precise value cannot be derived analytically because the need to perform the integral means that one must specify the density profile, and, in particular, the critical value of $\U$ for the onset of the instability varies between different types of star. Given the simple estimate here, it is unlikely that this instability is realised because such large values of $\U$ are already ruled out \cite{Sakstein:2015zoa}, and, even if they were not, values as large as this would make it very difficult to form realistic equilibrium stellar configurations \cite{Koyama:2015oma}.

\section{Semi-Analytic Stellar Models}\label{sec:semian}

In this section we numerically solve the LAWE for polytropic equations of state of the form
\begin{equation}\label{eq:poly}
P=K\rho^{\frac{n+1}{n}},
\end{equation}
where n is the polytropic index and K is a constant for stars of the same mass. Whilst not particularly realistic models of more complex stars, they can be used as simple models of homogeneous stars such as low mass main-sequence and dwarf stars. They have the added advantage that non-gravitational effects such as nuclear burning are decoupled from the hydrostatic equilibrium equation, allowing one to study the new effects of modified gravity in isolation. For this reason, their use has been particularly fruitful in previous modified gravity studies \cite{Davis:2011qf,Sakstein:2013pda,Koyama:2015oma,Sakstein:2015zoa,Sakstein:2015aac}. We first review the pertinent features of polytropes in beyond Horndeski theories (more detail and precise derivations can be found in \cite{Koyama:2015oma}) before solving the LAWE for several physically motivated models. 

\subsection{Polytropic Stars}

In GR, polytropic equations of state lead to a homology symmetry of the stellar structure equations \cite{1987A&A...177..117H} and so one can formulate the equations in a completely dimensionless form using the variables defined via
\begin{equation}
r_{\rm c}^2=\frac{(n+1)P_c}{4\pi G_N\rho_c^2},\quad r=r_{\rm c}y,\quad\rho=\rho_c \theta(y)^n,\quad \textrm{and}\quad P=P_c \theta(y)^{n+1},
\end{equation}
where $\rho_c$ and $P_c$ are the central density and pressure respectively. Note that $P_c=K\rho_c^{\frac{n+1}{n}}$. Substituting these into the hydrostatic equilibrium equation \eqref{eq:hse} and using the continuity equation $\dd M/\dd r = 4\pi r^2\rho$ one finds the Lane-Emden equation
\begin{equation}
\frac{1}{y^2}\frac{\dd}{\dd y}\left[\left(1+\frac{n\U\theta^{n-1}}{4}\right)y^2\frac{\dd\theta}{\dd y}+\frac{\U}{2}\xi^3\theta^n\right]=-\theta^n,
\end{equation}
which is solved subject the boundary conditions that $\theta(0)=1$ so that $P(0)_c$ and $\theta'(0)=0$, which is a requirement of spherical symmetry. Defining $y\ccc$ as the coordinate where the pressure falls to zero i.e. $\theta(y\ccc)=0$, the radius of the star is $R=r\ccc y\ccc$ and the mass of the star is given by 
\begin{equation}
M=4\pi r\ccc^3\rho_c\omega_R\quad\textrm{where}\quad\omega_R=-\frac{1}{y^2}\left.\frac{\dd \theta}{\dd y}\right\vert_{y\ccc}.
\end{equation}

Written in Lane-Emden variables, the LAWE \eqref{eq:MLAWE} is
\begin{align}
&\left[ \goz \theta + \frac{(n+1)}{4} \Upsilon x^2 \theta^n\right] \frac{d^2\xi}{dx^2}+\left[\frac{4}{x}\left(\goz \theta + \frac{(n+1)}{4} \Upsilon x^2 \theta^n \right) +\goz (n+1) \frac{d\theta}{dx} \right.\nonumber\\&\left.+ \frac{(n+1)}{4}\Upsilon \left(2x\theta^n + x^2 2n \theta^ {(n-1)} \frac{d\theta}{dx}\right)\right] \frac{d\xi}{dx}+\left[\frac{1}{x}(3\goz-4)(n+1)\frac{d\theta}{dx}\right]\xi = -\tilde{\omega}^2 W(\xi,\Upsilon) \xi, 
\end{align}
with
\begin{equation}
W(x,\Upsilon)=1+\Upsilon \frac{2x\theta'\theta^n}{(x\theta^n -4\theta')^2},\quad\textrm{and}\quad \tilde{\omega}^2=\frac{(n+1)}{4\pi G \rho\ccc}\omega^2.\label{eq:omegaLE}
\end{equation}
We have solved this for the polytropic stellar models given in table \ref{tab:stars}. These specific models have been chosen because polytropic models are reasonable approximations and because such objects have previously been shown to be sensitive to beyond Horndeski theories.  

When comparing stars in different theories of gravity, one always has the ambiguity in which properties are kept fixed. For example, one could fix the mass, in which case the central density and the radius will vary. In what follows we keep the mass fixed using the formula
\begin{equation}\label{eq:MR}
\frac{M}{R^3}=\frac{4\pi\rho\ccc\omega_R}{y\ccc^3}.
\end{equation}
Our procedure is the following: we first solve the Lane-Emden equation given a specific value of $n$ for GR, which gives us $\omega_R$ and $y\ccc$. Next, using the formula \eqref{eq:MR} we extract the value of $\rho\ccc$ given the stellar masses and radii in table \ref{tab:stars}. We then solve the Lane-Emden equation for various values of $\U$ and extract the requisite value of $\rho\ccc(\U)$ such that $M/R^3$ corresponds to the stars listed in table \ref{tab:stars}\footnote{The keen-eyed reader will have noticed that we have not needed to calculate the constant $K$ during this calculation. This is due to the homology symmetry mentioned above. This symmetry ensures that any star with mass $M$ and radius $R$ has the ratio $M/R^3$ fixed by $\rho\ccc$ at fixed $n$ and $\U$. One could use the individual values of $M$ and $R$ to find the value of $K$ but this is unnecessary for pulsations since the physical frequency depends on $\rho\ccc$ only (see equation \eqref{eq:omegaLE}). Said another way, the period of pulsation depends only on $\rho\ccc$ or, equivalently, $M/R^3$ at fixed $n$ and not the individual values of $M$ and $R$. This is in contrast to more realistic stellar models where the value of $M$ and $R$ is set by the internal microphysics and results in non-homologous pressure and density profiles.}. 

\begin{figure}[h]
{\includegraphics[width=0.49\textwidth]{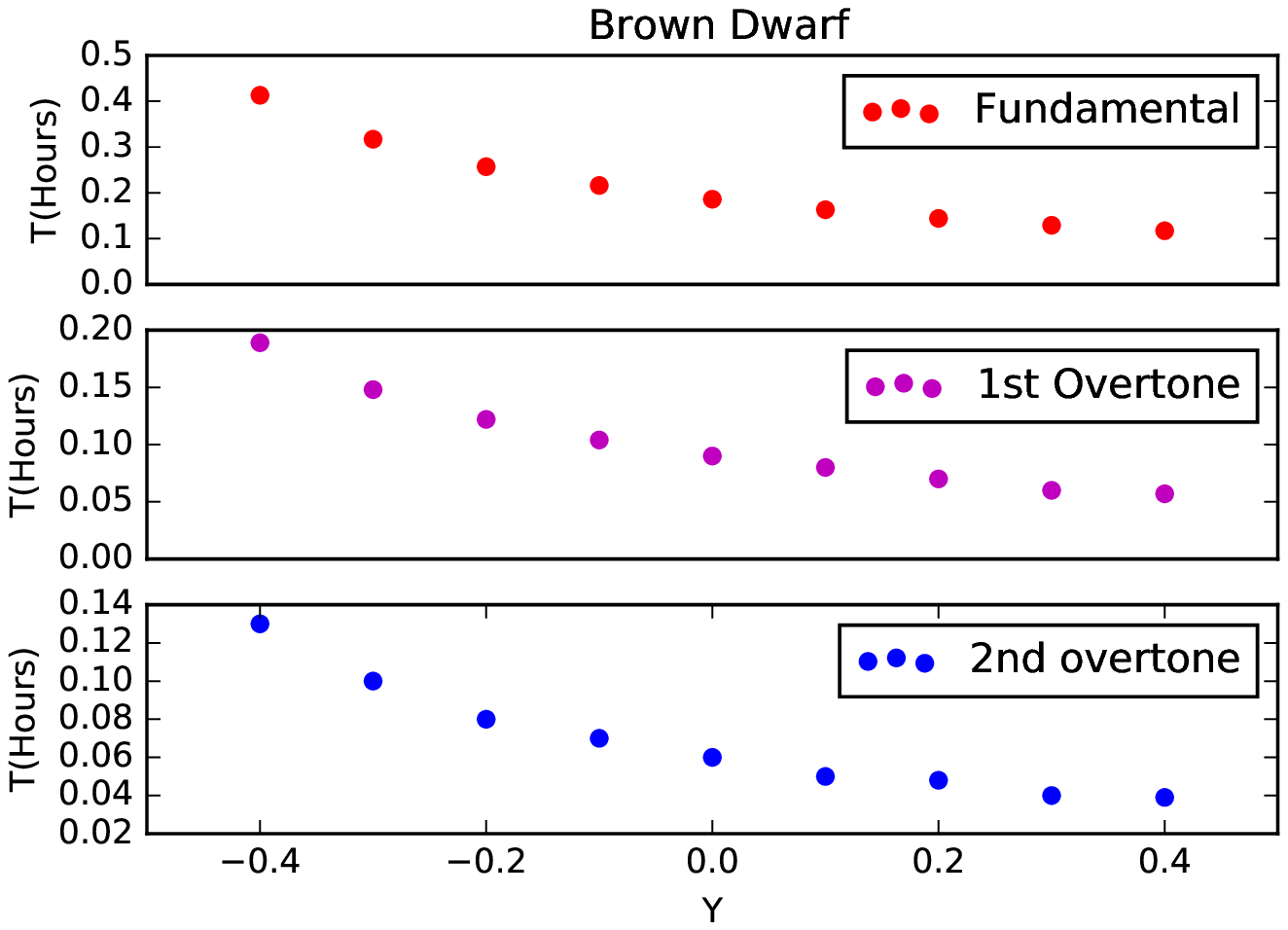}}
{\includegraphics[width=0.49\textwidth]{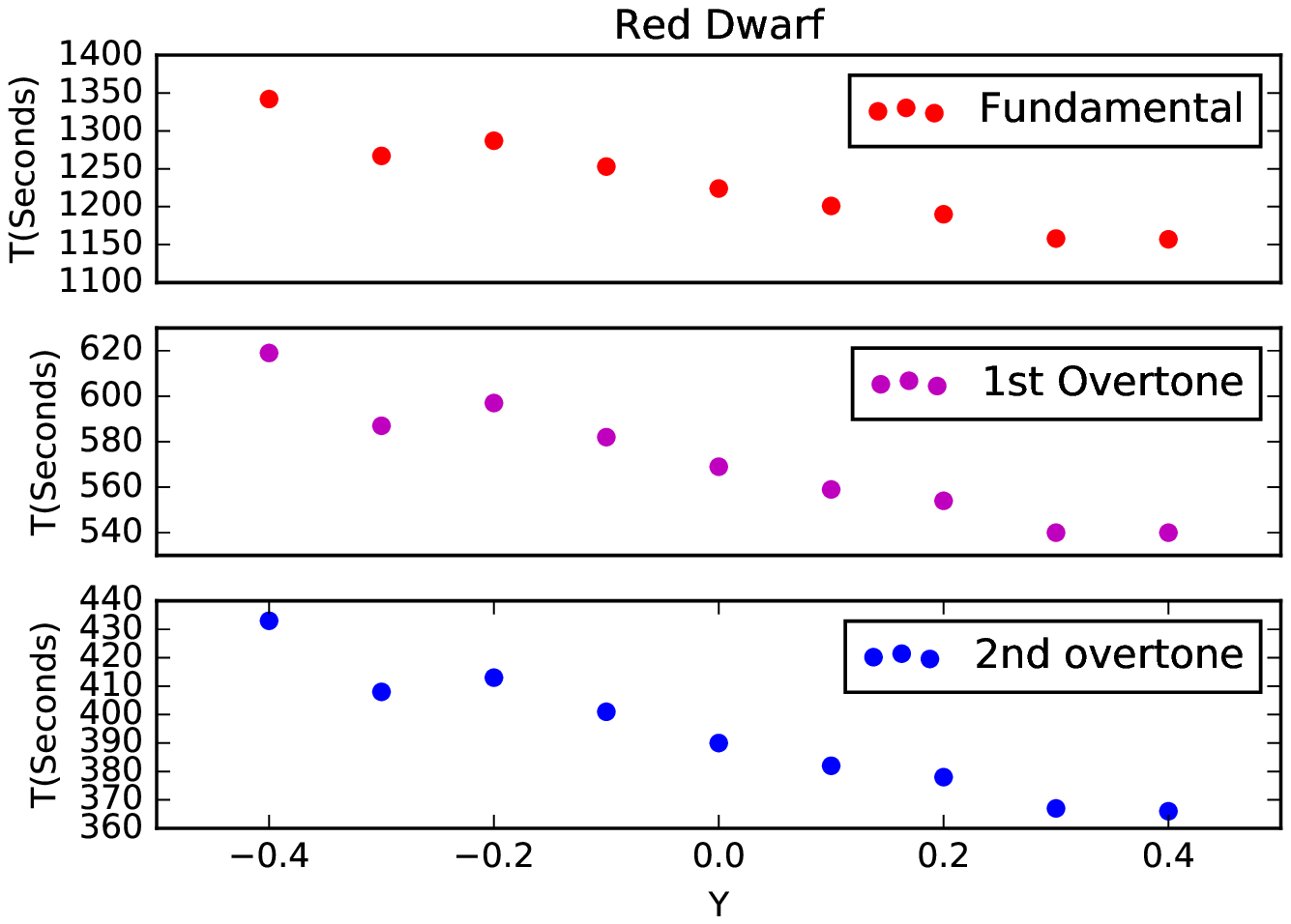}}
{\includegraphics[width=0.49\textwidth]{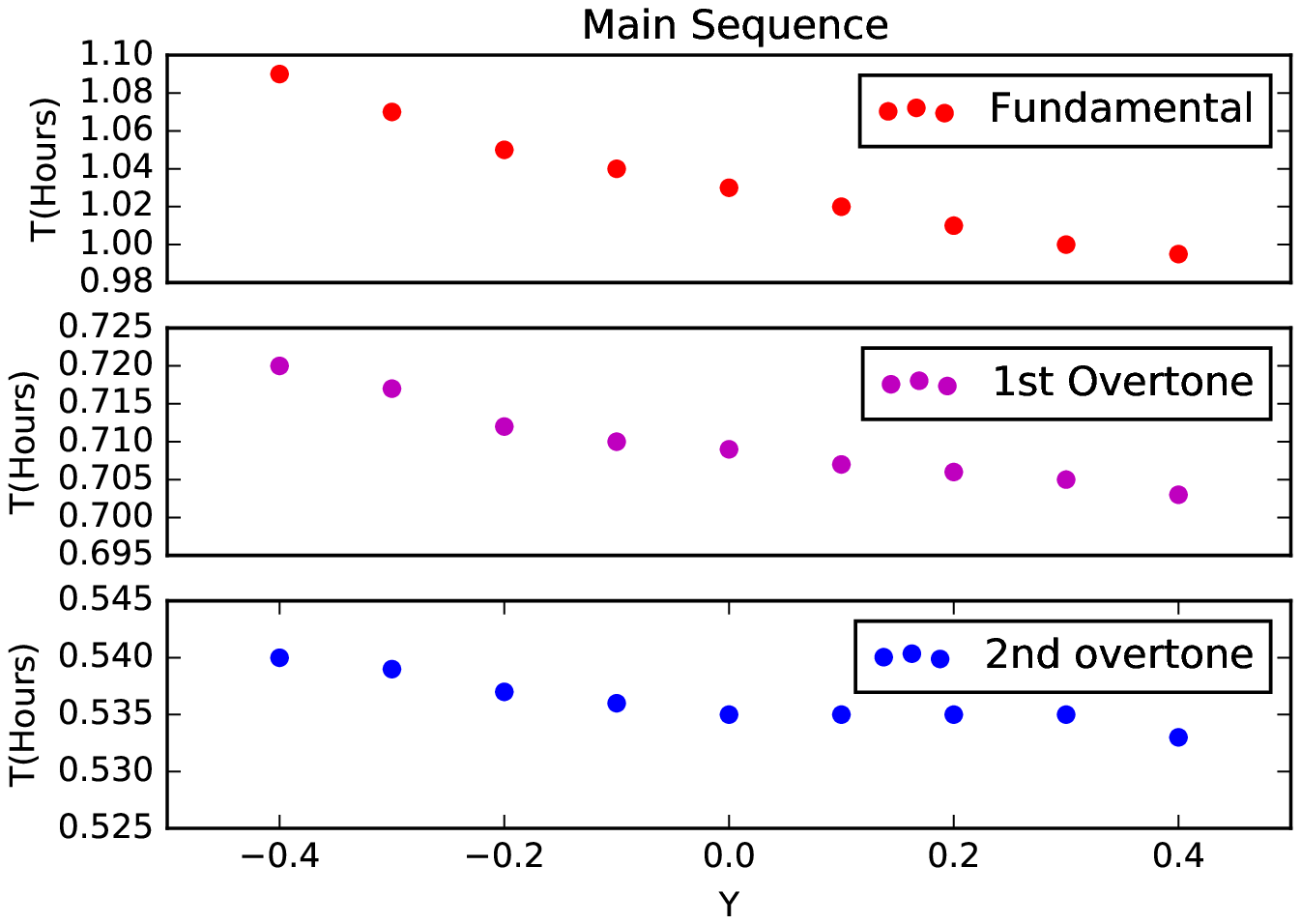}}
{\includegraphics[width=0.49\textwidth]{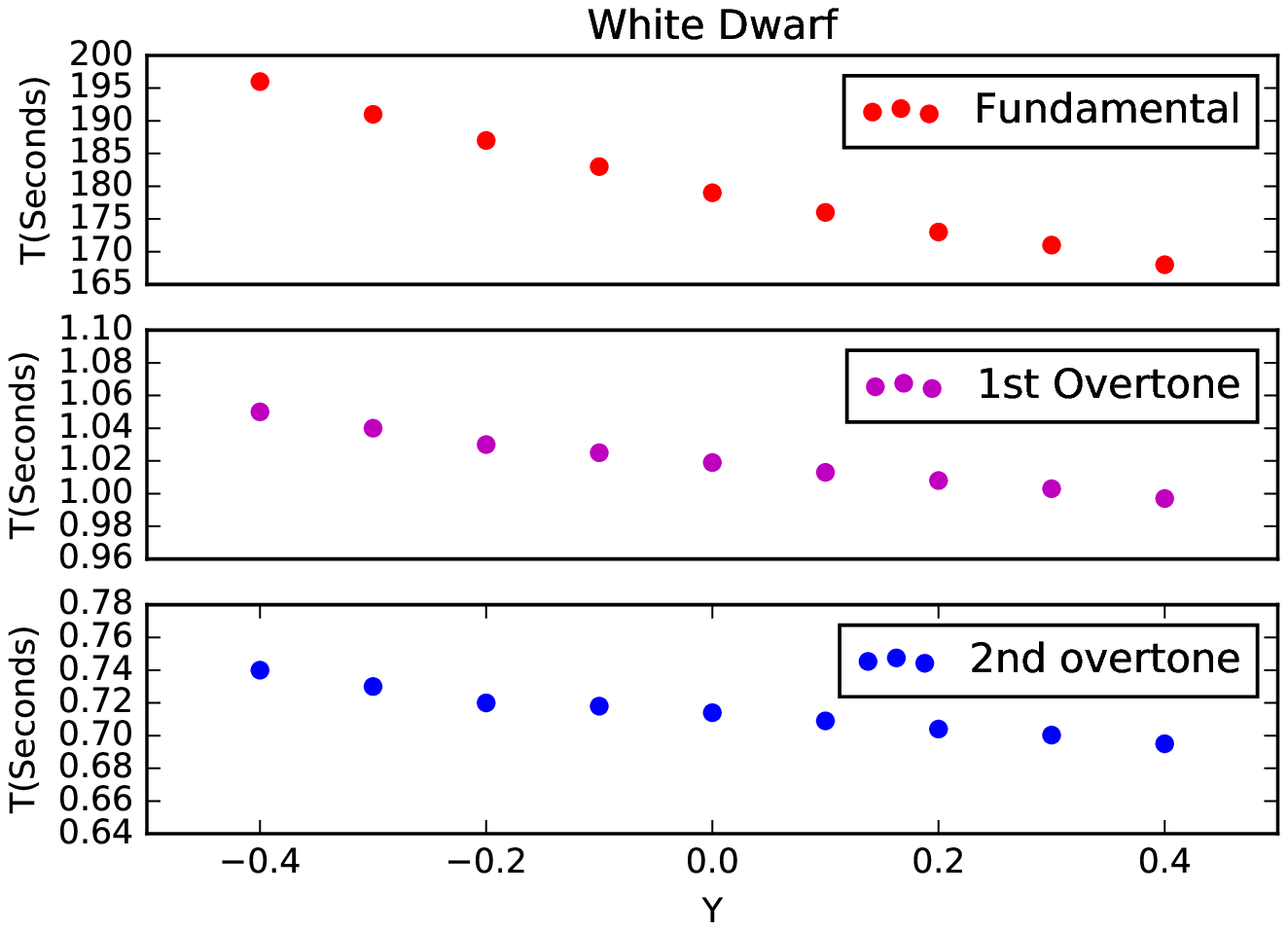}}
\caption{The period of the fundamental mode and the first and second overtones as a function of $\U$ for the stellar models given in table \ref{tab:stars}. The units for the period are given in the figures.}\label{fig:LEfreqs}
\end{figure}

\begin{figure}[h]\centering
{\includegraphics[width=0.85\textwidth]{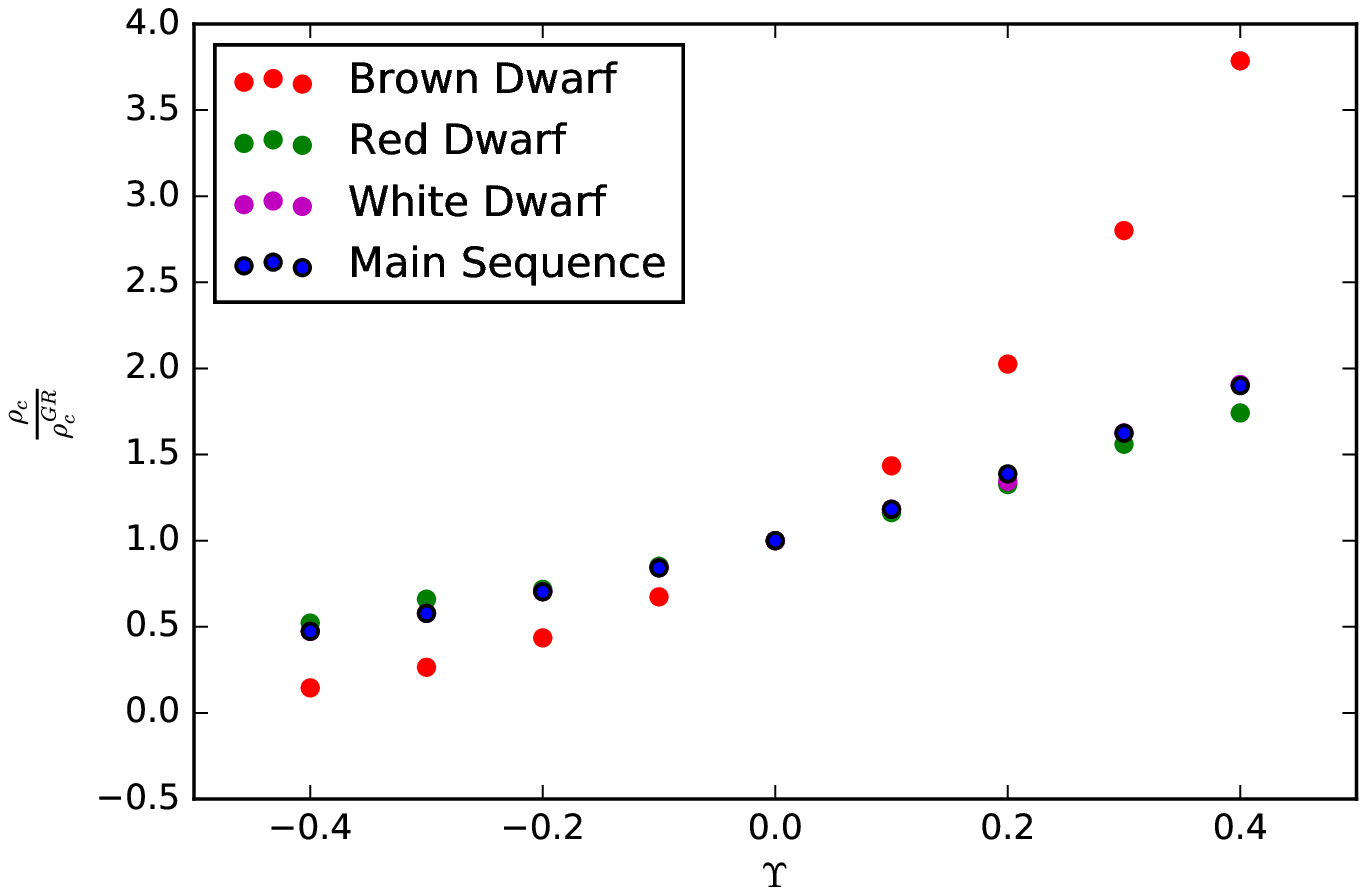}}
\caption{The ratio of the central density normalised to the GR value as a function of $\U$ for the stellar models given in table \ref{tab:stars}.}\label{fig:rhoc}
\end{figure}

The period of the fundamental mode and first and second overtones are shown in figure \ref{fig:LEfreqs}. One can discern the general trend that the changes due to modified gravity are more pronounced in the lower harmonics. The main-sequence, white dwarf, and red dwarf models show little change due to modified gravity (minutes on the hour or seconds on the minute) but the brown dwarf models exhibit significant changes so that $\delta T/T\sim 1$ and it may be possible to use these objects as probes. To understand why this is the case, we plot the central density (normalised to the GR value) as a function of $\U$ in figure \ref{fig:rhoc} since the central density affects the oscillation frequency via equation \eqref{eq:omegaLE}. One can see that the ratio varies by a factor of $\sim2$ for all models with the exception of brown dwarfs, where the variation is more dramatic. The other three models have relatively soft equations of state $P\sim\rho^{4/3}$, $\rho^{5/3}$ whereas the brown dwarf model's equation of state is much stiffer ($P\sim\rho^2$). This suggests that objects such as neutron stars, which typically require stiff equations of state to match observations, may be good probes of this theory, although such an investigation would require a fully relativistic treatment, which is well beyond the scope of this work. Interestingly, stellar evolution models suggested that young deuterium burning brown dwarfs are pulsationally unstable \cite{Palla:2005pr} but a subsequent targeted study detected no such instability \cite{2014ApJ...796..129C}. The data from this study may be useful for testing beyond Horndeski theories, although a more detailed modelling would be required to produce theoretical predictions that are precise enough to compare with data.

\begin{table}
\begin{center}
	\begin{tabular}{ c| c| c |c }
	Star  & $n$ & Mass ($M_{\odot}$)  &  Radius ($R_{\odot}$) \\ [0.5ex]
	\hline 
	 Brown Dwarf & 1 & 0.06 & 0.1 \\ [1ex]
	Red Dwarf & 1.5 & 0.2 & 0.2 \\ [1ex]
	Main-Sequence & 3 & 1 & 1 \\ [1ex]
	White Dwarf & 3   &  1.35  &  0.005
\end{tabular}\end{center}\caption{The polytropic stellar models investigated in this section.}\label{tab:stars}\end{table}

\section{Numerical Models}\label{sec:num}

In order to go beyond simple stellar models and include the essential microphysics that is missing in Lane-Emden models we require numerical simulations of the equilibrium stellar structure. In a previous paper \cite{Koyama:2015oma}, the stellar structure code MESA \cite{Paxton:2010ji,Paxton:2013pj,Paxton:2015jva} was modified to include the effects of beyond Horndeski theories. In what follows, we will solve the LAWE \eqref{eq:MLAWE} for two realistic stellar models:
\begin{enumerate}
\item Solar models: We evolve a star of $1 M_\odot$ with metallicity $Z=0.02$ from the pre-main-sequence until solar age and use the resulting model as input for the LAWE.
\item Cepheid models: We evolve a star of mass $4 M_\odot$ and metallicity $Z=0.004$ from the pre-main-sequence until its blue loop first crosses the blue edge of the instability strip given by $\log L = 4.2 - 46 (\log T_{\rm eff} - 3.8)$ \cite{Alibert:1999an}\footnote{Note that this has been found assuming GR. Finding the shift due to modified gravity would require non-adiabatic models, which is beyond the scope of this work. Any such shift is expected to be small since the mechanism driving the pulsations is due to non-gravitational physics.}. This model is used a input for the LAWE.
\end{enumerate}
The first of these models is by no means a realistic model of our Sun but serves as a useful proxy for understanding how the radial modes of Sun-like objects respond to changing the theory of gravity from GR to beyond Horndeski\footnote{Typically, a solar model is found by calibrating the model to match the observed age, luminosity, radius, and sound speed of the Sun by tuning free parameters controlling the microphysics (mixing length etc.), as well as the parameters controlling processes such as overshooting. It would be very difficult to disentangle the effects of modified gravity from that of changing these free parameters and so we do not attempt such a comparison here.}. The second set of models, Cepheid stars, are of particular interest since the low-order radial modes are observable via the period-luminosity relation, and because such objects have previously been incredibly successful at constraining alternative theories of gravity \cite{Jain:2012tn,Sakstein:2013pda,Sakstein:2014nfa}. 

\begin{figure}[h]
{\includegraphics[width=0.49\textwidth]{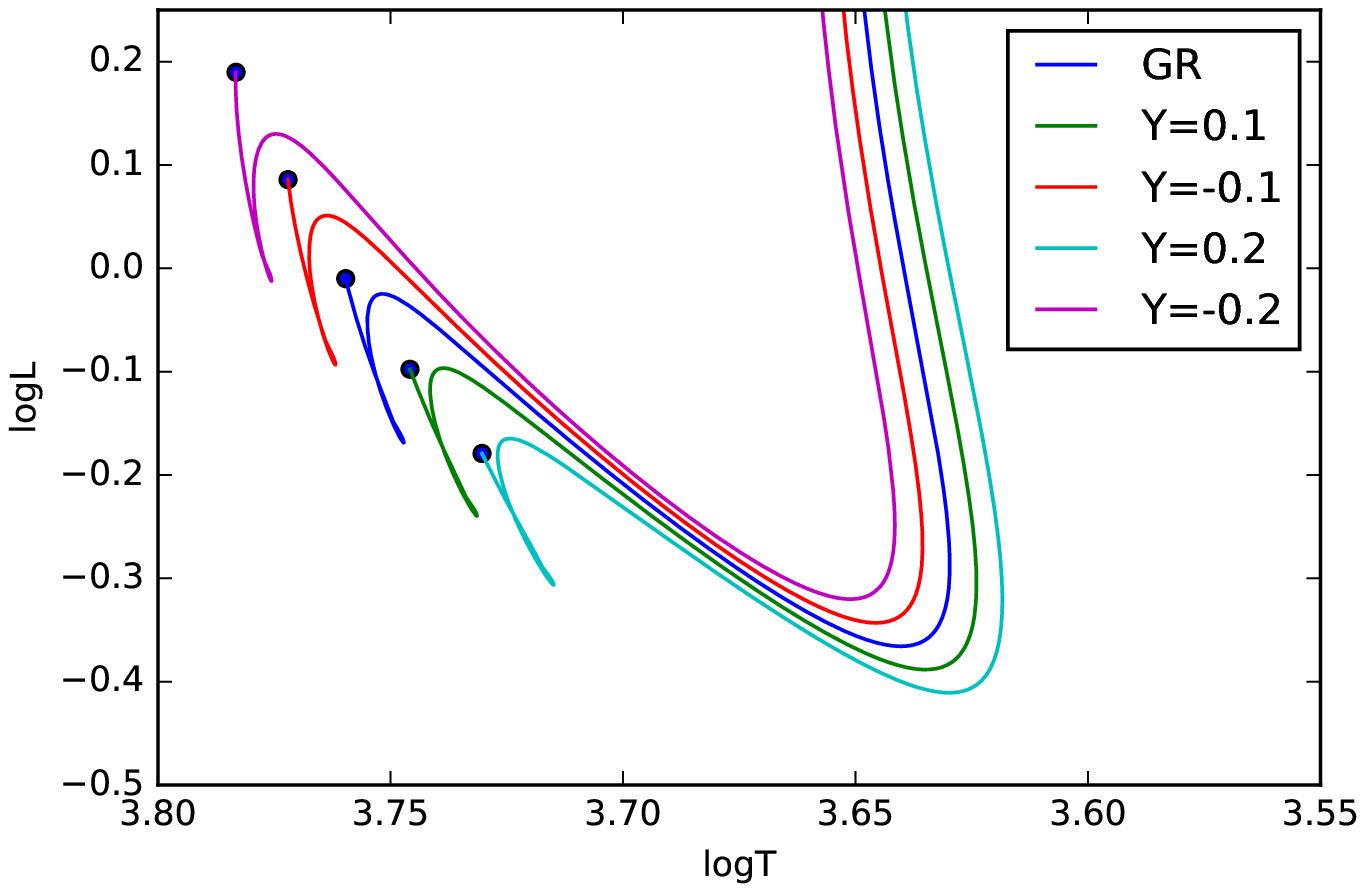}}
{\includegraphics[width=0.49\textwidth]{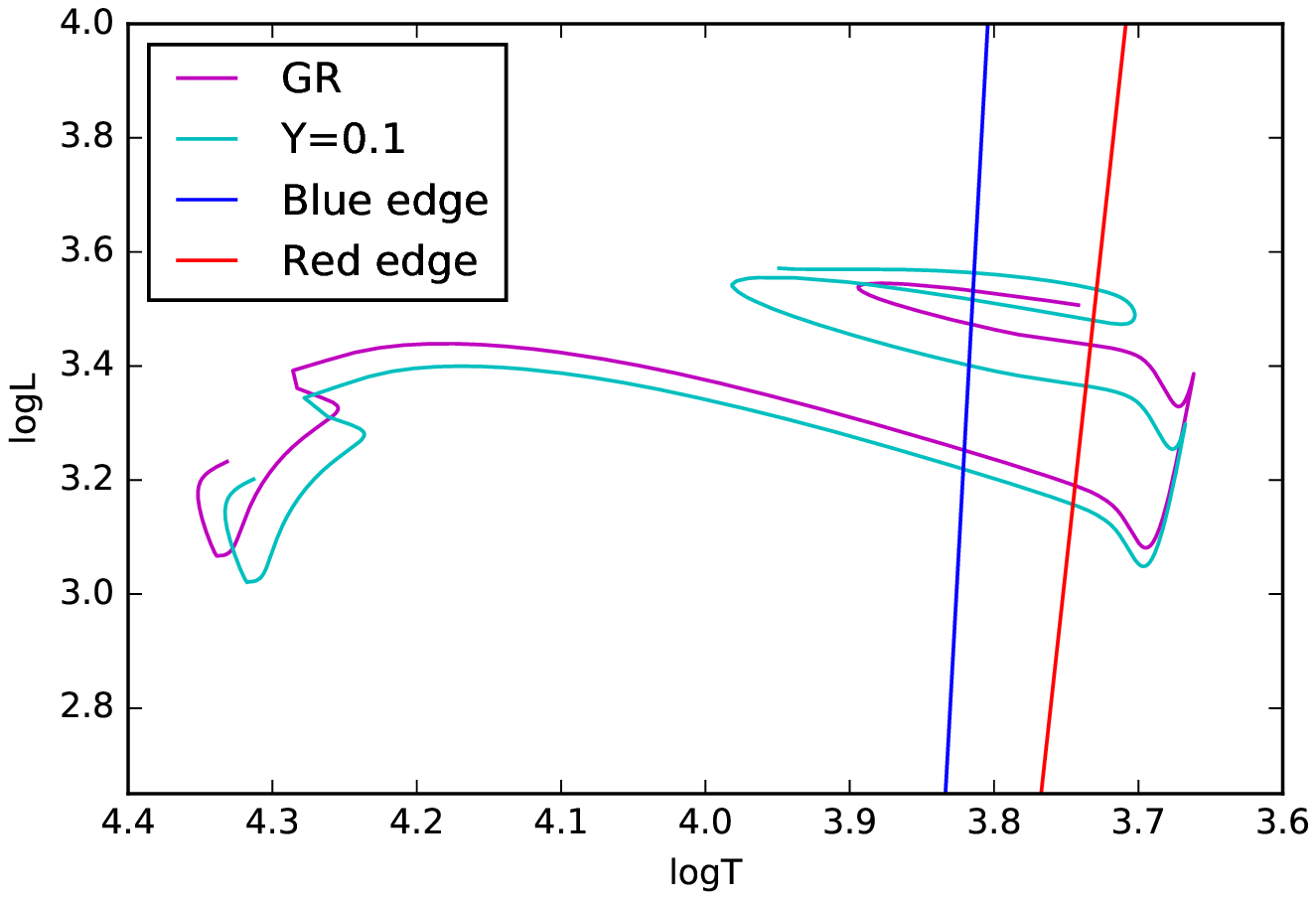}}
\caption{The HR tracks for the two models investigated in this section. \emph{Left Panel}: The HR tracks for solar models ($M=M_\odot$, $Z=0.02$). The values of $\U$ are indicated by the legend in the figure and the models we use to solve the LAWE (i.e. those where the age of star is equal to the Sun's age) are shown by the black points. \emph{Right Panel}: The HR tracks for Cepheid models. Note that only $\U=0.1$ is shown for clarity.}\label{fig:HRs}
\end{figure}

\begin{figure}[h]
{\includegraphics[width=0.49\textwidth]{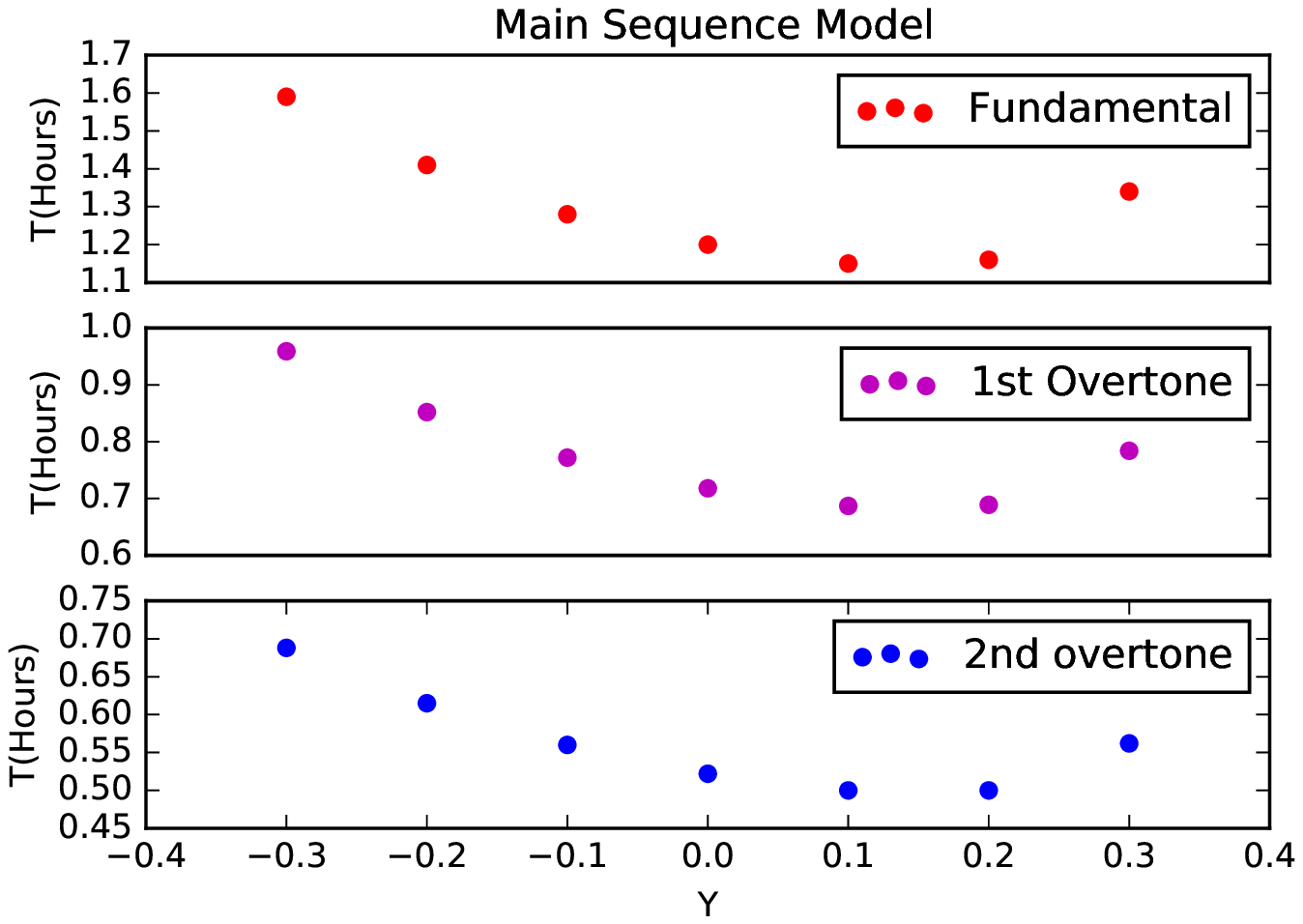}}
{\includegraphics[width=0.49\textwidth]{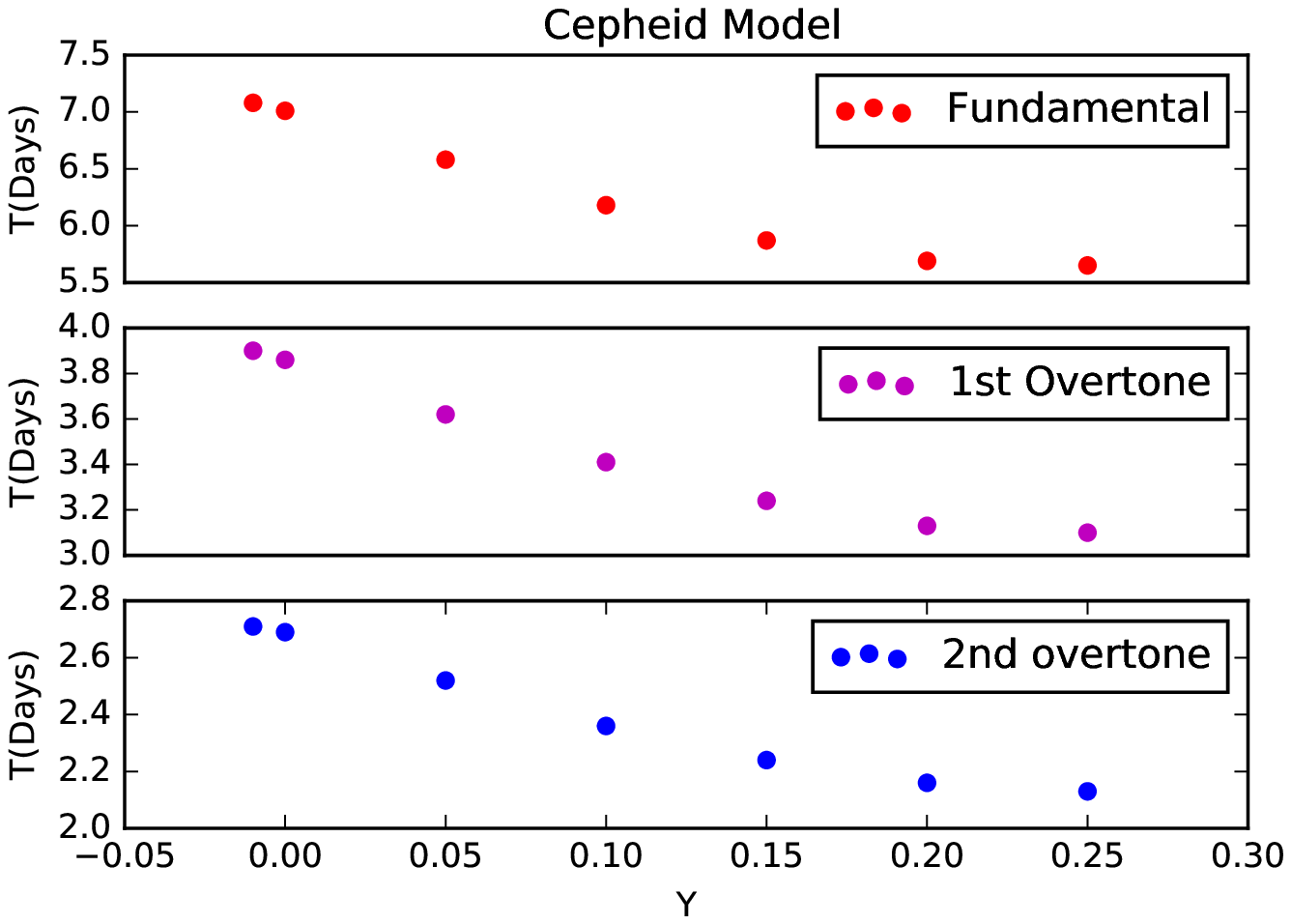}}
\caption{The period of the fundamental mode and the first and second overtones for solar models (left panel) and Cepheid models (right panel) as a function of $\U$. Note that the units are given on the axes.}\label{fig:numfreqs}
\end{figure}

The Hertzprung-Russell (HR) tracks for these models are shown in figure \ref{fig:HRs} and the resultant time-periods are shown in figure \ref{fig:numfreqs}. Note that the upturn in the periods of solar models is not unexpected; the weight function $W(r,\Upsilon)$ is not symmetric about $\U=0$ and so there is no a priori reason why the GR frequency should be a stationary point. In both cases, the changes due to modified gravity are largest for the fundamental mode and decrease with increasing overtone. The changes in the solar model amount to minutes on the hour whereas the changes in the Cepheid model are of order days, making them very sensitive objects. This is partly because the solar HR tracks are only slightly modified by beyond Horndeski theories whereas the Cepheid tracks, and consequentially the intercept with the blue edge of the instability strip, are drastically altered. 

Cepheids have previously been used to constrain chameleon models \cite{Jain:2012tn} by comparing the distance to unscreened galaxies measured using the period-luminosity relation with the distance measured using the tip of the red giant branch, which is largely insensitive to modified gravity. A similar test could be performed for beyond Horndeski theories since previous studies have found that the tip of the red giant branch is not sensitive to beyond Horndeski modifications of gravity \cite{Koyama:2015oma}. Indeed, such a test would be easier for this class of theories since all galaxies are unscreened. Another possibility is to use systems such as OGLE-LMC-CEP-0227 \cite{Pietrzynski:2010tr,Marconi:2013tta} (see \cite{Sakstein:2014nfa} for a discussion of testing gravity with such a system). This system contains a Cepheid in an eclipsing binary so that one can measure the Cepehid's mass using both the eclipsing binary technique and pulsation models; both mass estimates agree at the $1\sigma$ level. The orbital dynamics in beyond Horndeski theories are identical to GR whereas the pulsation period is not. Therefore, beyond Horndeski theories predict that the two mass estimates should not agree (or rather, should agree if one uses beyond Horndeski pulsation models) so that the agreement in this system can place new constraints on $\U$.


\section{Discussion and Conclusions}\label{sec:concs}

In this work we have derived the wave equation governing small time-dependent radial perturbations of stars in beyond Horndeski theories of gravity. The eigenvalues of this equation give the frequency of stellar pulsations and we have calculated these for various semi-analytic (polytropic) and realistic numerical stellar models in a first attempt to survey the stellar zoo and identify the best candidate objects for testing these theories. We have also analytically identified a new potential instability whereby there are growing mode solutions of the LAWE when $\U\gsim49/6$, although values this large are already ruled out observationally \cite{Sakstein:2015zoa}. 

Our analysis has identified two potential candidates which show significant deviations from GR: brown dwarf stars and Cepheids, both of which exhibit large deviations so that $\delta T/T\sim\oo(1)$. We have discussed how each in turn may be used as a probe of modified gravity. Brown dwarfs have been surveyed extensively and there is some data pertaining to their variability, although a more detailed theoretical modelling of the interior physics and a thorough investigation of degeneracies with quantities such as the mass are required. Cepheids are a more likely candidate due to the abundance of data and their history as probes of modified gravity. Indeed, we have discussed two possible methods---comparing distance estimates and mass estimates in eclipsing binaries---that could be used to constrain $\U$.

It would be interesting to study non-radial modes, although, as we have discussed above, the derivation of the wave equations would be very difficult due to the need to start from a fully relativistic description and account for the cosmological time-variation of the scalar. Whilst not an a priori impossible task, this clearly lies beyond the scope of this work. A simpler extension of this work would be to include non-adiabatic effects, which is tantamount to sourcing the LAWE with a driving term and therefore does not require additional input from gravitational physics. Solving the non-adiabatic problem would allow one to predict the amplitude of the pulsations as well as the location of the instability strip, both of which may serve as additional tests of gravity. Finally, we have noted that objects with stiffer equations of state tend to be more sensitive to modifications of gravity. This suggests that a study of pulsating compact objects such as neutron stars may be fruitful, although degeneracies with the equation of state may make it difficult to place any constraints. 

\section*{Acknowledgements}

We are gratelful to Thomas Tram for several enlightening disucssions. KK is supported by the UK Science and Technologies Facilities Council grants  ST/N000668/1 and the European Research Council through grant 646702 (CosTesGrav). We are grateful to the SEPnet summer placement scheme for its support in organising the placement in which MK-A has participated, and to the University of Portsmouth for hosting MK-A during the time this work was carried out.

\appendix

\section{Model}\label{sec:model}

The specific model we study in this work is the model studied by \cite{Babichev:2016jom}, which is a covariantisation of the quartic galileon\footnote{Note that we work with the $(-,\,+,\,+,\,+)$ signature for the metric so that $k_2<0$ corresponds to a healthy ghost-free (correct sign) canonical kinetic term. We will see shortly (equation \eqref{eq:f4k2rel}) that our model statisfies this condition. }
\begin{equation}
\label{action}
S=\int d^4x\,  \sqrt{-g}\, \left[\mpl^2\left(\frac{R}{2}-\Lambda\right)-k_2X +f_4\lag_{4,{\rm bH}}\right]\,,
\end{equation}
with
\begin{align}
\mathcal{L}_{4,{\rm bH}}&=-X\left[(\Box\phi)^2-(\phi_{\nm})^2\right]+2\phi^\mu\phi^\nu\left[\phi_\nm\Box\phi-\phi_{\mu\sigma}\phi^{\sigma}_{\,\,\nu}\right],
\end{align}
with X=$\phi_{\mu}\phi^{\mu}$ and where $\Lambda$ is a cosmological constant\footnote{This is required in order to give Vainshtein breaking \cite{Babichev:2016jom}.}. This is the simplest subset of beyond Horndeski theories that admits Vainshtein breaking inside astrophysical bodies\footnote{There are more complicated models where cosmological constant is not strictly necessary because there are other, more general models that give Vainshtein breaking without it \cite{Sakstein:2016oel}. These models have more complicated kinetic terms so that $k_2$ is now a function of $X$ but give identical equations of motion in the sub-Horizon limit because $k_2(X)\propto H^2$. The deviations from GR are still governed by a single parameter $\Upsilon$; the only difference is how it is related to the underlying parameters appearing in the Lagrangian. For this reason, the LAWE we have used in this work applies equally to these theories as well. We have chosen to present only the model \eqref{action} here for the sake of simplicity.  }. Cosmological de-Sitter solutions exist at late-times so that the line-element is 
\begin{equation}
\label{metric_dS}
\dd s^2 = -\dd \tau^2 + e^{2H\tau} \left(\dd\rpp^2 + \rpp^2\dd\Omega_2^2 \right) 
\end{equation}
in the Freidmann-Robertson-Walker (FRW) slicing. Here,
\begin{equation}
\phi=\pdot \tau,\quad
k_2=-2\frac{\mpl^2H^2}{v_0^2}\left(1-\sigma^2\right),\quad \textrm{and}\quad
f_4=\frac{\mpl^2}{6v_0^4}\left(1-\s^2\right)\label{eq:f4k2rel},
\end{equation}
where $\s^2\equiv \Lambda/(3 \mpl^2 H^2)$.

\section{Derivation of the Wave Equation}\label{sec:deriv}

In this appendix we formally derive the linear adiabatic wave equation for the model presented in Appendix \ref{sec:model}. This is by no means a complete derivation and we will often refer to other works on the equilibrium structure of relativistic stars \cite{Babichev:2016jom} in these theories and use well-established results of perturbation theory applied to astrophysical fluids \cite{cox1980theory,thompson2006introduction} without proof.

In order to derive the wave equation we need to calculate the Newtonian potential $\Phi$, defined in the Newtonian gauge via
\begin{equation}
\dd s^2 = -(1+2\Phi)\dd t^2 + (1+2\Psi)\delta_{ij}\dd x^i\dd x^j
\end{equation}
so that we may close equation \eqref{eq:eqhse}. For the purposes of our calculation, it is more convenient to work in Schwarzchild-de Sitter coordinates\footnote{Note that since we only need the $00$-component to close the system of equations the coordinates chosen for the spatial part of the metric are irrelevant and so it does not matter whether we use Schwarzchild or isotropic coordinates. The transformation between the two does not change the definition of $r$ at Newtonian order.} defined by
\begin{equation}
\dd s^2 = -e^{\nu(r,t)}\dd t^2 +e^{\lambda(r,t)}\dd r^2 + r^2\dd\Omega_2^2,
\end{equation}
where we take
\begin{align}
 \nu(r,t)&=\ln\left(1-H^2r^2\right)+\delta\nu_0(r) + \dn_1(r)e^{i\omega t}, \label{eq:nusubh} \\
 \lambda(r)&=-\ln\left(1-H^2r^2\right)+\dl_0(r) + \dn_1(r)e^{i\omega t} \label{eq:lamsubh} \quad \textrm{and} \\
 \phi(r,t) &= v_0t+\frac{v_0}{2H}\ln\left(1-H^2r^2\right)+\varphi_0(r)+\varphi_1(r)e^{i\omega t}.\label{eq:phisubh}
\end{align}
Here, subscript zeros refer to the equilibrium structure of the star and subscript ones refer to linear time-dependent perturbations. When these are zero one finds the de Sitter metric in Schwarzchild coordinates (see \cite{Babichev:2016jom}) and so one should view this expansion as first taking the de Sitter solution in the Schwarzchild slicing, adding a static spherically symmetric object which sources the quantities with zero subscripts and then making small time-dependent radial perturbations that source the quantities with subscript ones. The equilibrium structure in the sub-Horizon weak-field limit was already found by \cite{Babichev:2016jom} who report
\begin{align}
\dn_0'(r)&=\frac{2\GN M(r)}{r^2}+\frac{\U\GN}{2}M''(r),\quad\dl_0(r)=\frac{2\GN M(r)}{r^2}-\frac{5\U\GN}{2}M'(r),\quad\textrm{and}\nonumber\\\frac{\varphi_0(r)}{\pdot}&=-\sqrt{\frac{2\GN M(r)}{r}+\frac{1}{2}\GN M'(r)},\label{eq:equil}
\end{align}
where
\begin{align}
\GN=\frac{3}{8\pi(5\s^2-2)\mpl^2}\quad\textrm{and}\quad
\U=-\frac{1}{3}\left(1-\s^2\right).
\end{align}
These are quantities that are found by solving the equilibrium stellar structure equations and hence appear as pre-known functions when solving the LAWE. 

Next, we want to find the equations governing the time-evolution of the perturbations of these quantities in the weak-field limit, which, as discussed in \cite{Babichev:2016jom}, corresponds to scalings of the form $\dn_0\sim\dl_0\sim GM/R$ and $\varphi_0\sim \sqrt{GM/R}$. We thus take $\dn_1\sim\dl_1\sim (\delta r/r)GM/R$ and $\varphi_1\sim (\delta r/r)\sqrt{GM/R}$ so that these quantities are both 1PN in the PPN counting scheme and small perturbations. Additionally, $\omega^2\sim GM/R^3$. With this in hand, the $00$-component of the tensor equation of motion yields the equation
\begin{equation}
\frac{\dl_1}{1+5\U}+\frac{r\dl_1'}{1+5\U}
+10\frac{\varphi_0'\varphi_1'}{\pdot^2(1+5\U)}
+10\frac{\U r\varphi_1'\varphi_0''}{\pdot^2(1+5\U)}+10\frac{\U r\varphi_0'\varphi_1''}{\pdot^2(1+5\U)}=8\pi \GN r^2\rho_1(r),
\end{equation}
where we have expanded the Eulerian energy density as $\rho(r)=\rho_0(r)+\rho_1(r)e^{i\omega t}$. This equation can be integrated once with the aid of the first-order fluid identity
\begin{equation}\label{eq:rho1}
\rho_1=-\frac{1}{r^2}\frac{\dd}{\dd r}(r^2\rho_0\delta r),
\end{equation}
which comes from perturbing the continuity equation $\nabla_\mu T^{\mu\nu}=0$ and is therefore not altered in beyond Horndeski theories, to find
\begin{equation}\label{eq:lam}
\dl_1=-8\pi(1+5\U)\GN r\rho_0\delta r-10\frac{\U\varphi_0'\varphi_1'}{\pdot^2}.
\end{equation}
This may be substituted into the $rr$-component of the tensor equation to find a formula for $\dn_1'(r)$:
\begin{equation}\label{eq:nu}
\dn_1'(r)=-8\pi(1+5\U)\GN\delta r -2i\omega\frac{\Upsilon\varphi_1'}{\pdot}-8\frac{\U\varphi_0'\varphi_1'}{\pdot^2 r}+2\frac{\U\varphi_0''\varphi_1'}{\pdot^2}+2\frac{\U\varphi_0'\varphi_1''}{\pdot^2}.
\end{equation}
Note that this depends on $\varphi_1$ but not $\dl_1$. The presence of terms of the form $i\omega f(\varphi_0,\varphi_1)$ are interesting because they may signal new instabilities that could drive stellar pulsations but, as we will see shortly, they are highly suppressed. One lesson to be learned from previous analyses is that the scalar equation $\nabla_\mu j^\mu=0$, where the current $j^\mu$ is given in \cite{Babichev:2016jom}, must be expanded to $\mathcal{O}(GM/R)^{3/2}$. At the level of perturbations, this yields, after a trivial integration and upon substituting \eqref{eq:lam} and \eqref{eq:nu},
\begin{equation}
\frac{\varphi_1'}{\pdot}=2\pi i\omega\frac{r^3\rho_0\delta r}{4M+rM'}+6\sqrt{2}\pi\frac{\sqrt{\GN}r^{\frac{3}{2}}\rho_0\delta r}{\sqrt{4M+rM'}}+\sqrt{2}\pi\frac{\sqrt{\GN}r^{\frac{5}{2}}(\rho_0\delta r'+\rho_0'\delta r)}{\sqrt{4M+rM'}}.
\end{equation}
The term proportional to $i\omega$ is suppressed because it scales as $\sqrt{GM/R}(\delta r/r)(r/R)^3$, which is suppressed relative to the other terms by a factor of $(r/R)^3$ except in the outer layers of the star. Since these have little to no influence on the oscillation of the star we can safely ignore this term from here on. In this case, one can substitute the solution for $\varphi_1$ into equation \eqref{eq:nu} to find 
\begin{align}\label{eq:nu1}
\delta\nu_1'&=-\frac{2GM'}{r^2}\delta r+\frac{\Upsilon\GN(M'''\delta r-M'\delta r''-2M'')}{2}+\frac{4\U\omega^2rMM'\delta r}{(4M+rM')^2},
\end{align}
where the first term is identical to the GR result.

Using the fact that $\dd\Phi/\dd r = \dn_1'/2$ we can finally close the momentum equation. Perturbing the displacement, the pressure, and the Newtonian potential as 
\begin{align}
\frac{\delta r}{r}& = \xi(r)e^{i\omega t},\label{eq:drexp}\\
P(r,t)&=P_0(r)+P_1(r)e^{i\omega t},\quad\textrm{and}\\
\Phi'(r,t)&= \frac{\dn_0'}{2}+\frac{\dn_1'}{2}e^{i\omega t},
\end{align}
where $P_0$ is the solution of the hydrostatic equilibrium equation and $\Phi_0'$ is given by equation \eqref{eq:mod1}, the momentum equation becomes
\begin{equation}
\omega^2r\rho_0\xi(r)=\frac{1}{2}\dn_0\rho_1+\frac{1}{2}\rho_0\dn_1'+P_1'.
\end{equation}
We can eliminate the pressure using a standard result found by perturbing the equation of state and assuming adiabaticity:
\begin{equation}\label{eq:p1}
P_1(r)=\frac{\goz P_0}{\rho_0}\left[\rho_1+r\rho_0'\xi\right]-rP_0'\xi.
\end{equation}
Substituting equations \eqref{eq:hse}, \eqref{eq:equil}, \eqref{eq:rho1}, \eqref{eq:nu1}, \eqref{eq:drexp}, and \eqref{eq:p1} into the momentum equation one finds a wave equation of the form 
\begin{equation}\label{eq:int}
P(r)\xi''+Q(r)\xi'+R(x)\xi + \omega^2 S(r)\xi = 0,
\end{equation}
where
\begin{align}
T(r)&=64r^4\rho_0\left(\goz P_0+\pi\U\GN r^2\rho_0^2\right)\left(P_0'+\pi\U\GN r\rho_0\left[(2\U-1)\rho_0+\U r\rho_0'\right]\right)^2,\\
Q(r)&=64r^3\rho_0\left(P_0'+\pi\GN r\rho_0\left[(2\U-1)\rho_0+\U r \rho0'\right]^2\left[\goz\left(4P_0+rP_0'\right)\right.\right.\nonumber\\&\left.\left.+r\left(P_0\goz'+2\pi\U\GN r\rho_0\left[3\rho_0+r\rho_0'\right]\right)\right]\right),\\
R(r) & =64r^3\rho_0\left(3P_0\goz'+(3\goz-4)P_0'\right)\left(P_0'+\pi\GN r\rho_0\left[(2\U-1)\rho_0+\U r\rho_0'\right]\right)^2,\\
S(r)& =32r^4\rho_0^2\left(2\GN^2\pi^2r^2\rho_0^4\left[1+\U(5\U-4)\right]+2P_0'^2+\pi^2\U\GN^2r^3\rho_0^3\rho_0'(9\U-4)\right.\nonumber\\&\left.+4\pi\U\GN r^2\rho_0P_0'\rho_0'+\pi\GN r\rho_0^2\left[(9\U-4)P_0'+2\pi\GN\U^2r^3\rho_0'^2\right]\right).
\end{align}
This is not of the Sturm-Liouville type but we can multiply by an integrating factor
\begin{equation}
\frac{1}{T(r)}e^{\int\frac{Q(r)}{T(r)}\dd r}=r^4\frac{\goz P_0+\pi\U\GN r^2\rho_0^2}{T(r)},
\end{equation}
which makes the $\xi''$ and $\xi'$ terms into an exact differential. Upon doing this, one finds precisely equation \eqref{eq:MLAWE}.

\bibliographystyle{jhep}
\bibliography{ref}
\end{document}